\documentclass[twocolumn,aps,showpacs,nofootinbib,epsfig]{revtex4}
\usepackage{epsfig}
\usepackage{amsmath}
\usepackage{graphics}
\newcommand{\be}{\begin{equation}}
\newcommand{\ee}{\end{equation}}
\newcommand{\bea}{\begin{eqnarray}}
\newcommand{\eea}{\end{eqnarray}}
\newcommand{\beas}{\begin{eqnarray*}}
\newcommand{\eeas}{\end{eqnarray*}}

\newcommand{\slsh}[1]{{\not \! #1}}

\newcommand{\kv}{\boldsymbol{\mathrm{k}}}
\newcommand{\kvp}{\boldsymbol{\mathrm{k}^{\prime}}}
\newcommand{\vv}{\boldsymbol{\mathrm{v}}}
\newcommand{\qv}{\boldsymbol{\mathrm{q}}}

\begin{document}
\title{Collisional parton energy loss in a finite size QCD medium revisited:
  Off mass-shell effects} 
\author{Alejandro Ayala$^\dagger$, J. Magnin$^{\ddagger}$, Luis Manuel
Monta\~no$^*$ and Eduardo Rojas$^\dagger$}

\affiliation{$^\dagger$Instituto de Ciencias Nucleares, Universidad
Nacional Aut\'onoma de M\'exico, Apartado Postal 70-543, M\'exico
Distrito Federal 04510, M\'exico.\\
$^\ddagger$Centro Brasileiro de Pesquisas Fisicas, Rua Dr. Xavier
Sigaud 150-Urca CEP 22290-180, Rio de Janeiro, Brazil.\\
$^*$Centro de Investigaci\'on y de Estudios Avanzados del IPN,
Apartado Postal 14-740, M\'exico Distrito Federal 07000, M\'exico.}

\begin{abstract}

We study the collisional energy loss mechanism for particles produced off
mass-shell in a finite size QCD medium. The off mass-shell effects introduced
are to consider particles produced in wave packets instead of plane waves and
the length scale associated to an in-medium particles' life-time. We
show that these effects reduce the energy loss as compared to the case when
the particles are described as freely propagating from the source. The
reduction of the energy loss is stronger as this scale becomes of the order or
smaller than the medium size. We discuss possible consequences of the result
on the description of the energy loss process in the parton recombination
scenario. 

\end{abstract}


\maketitle

\section{Introduction}\label{sec1}

The problem of collisional parton energy loss in a QCD medium has been
revived due to recent RHIC data on non-photonic single
electrons~\cite{RHIC} that are not well described within radiative energy
loss calculations. Collisional energy loss has been a subject of research 
from long ago~\cite{{Bjorken},{Gyulassy1},{Gyulassy2}}. During the 
pioneering years, an important question was to understand how to handle the
infrared singularities in perturbative calculations at finite
temperature. The advent of resummation techniques clarified this point and
allowed to reliably compute the energy loss of a heavy parton traversing an
infinie medium to lowest order in perturbation
theory~\cite{{Braaten1},{Braaten2}}. Short after, it was estimated
that radiative energy-loss in a finite size medium was a more important
mechanism to account for energy losses of energetic
partons~\cite{{Baier1},{Baier2},{Gyulassy3}}. Nonetheless, even more recent
studies~\cite{{Mustafa},{Dutt}} suggested that for a range of parameters
relevant to RHIC energies, radiative and collisional energy losses for heavy
quarks are of the same order of magnitude. These last calculations where done
for infinite QCD media. The outstanding question was whether collisional
energy loss for finite size media was also significant.  

In this context there where two results seemingly in
contradiction~\cite{Peigne, Djordjevic}. In Ref.~\cite{Peigne} a semiclassical
approach based on linear response theory computes the collisional energy loss
by means of the work done by the response chromoelectric field on the
color-charged heavy parton traversing the medium. The infinite medium limit of
this description agrees with the collisional energy loss result at high
temperature --up to color factors-- obtained from a perturbative approach
using HTL effective propagators~\cite{Braaten1}. The original claim that the
finite size medium induced energy loss is strongly suppressed compared to the
infinite medium case was later revised by properly subtracting the kinetic
energy associated to producing the particle within the medium~\cite{Gossiaux}.

On the other hand, in Ref.~\cite{Djordjevic} a lowest order perturbative
calculation using HTL propagators finds that finite size effects on the
collisional energy loss are not significantly suppressed as compared to the
infinite medium case. The formulation of the problem is based on the
assumption that the scattered particle originates within the medium but
otherwise is produced on mass-shell.

However, when particles are emitted by sources lasting a finite amount of time
they are not necessarily produced on their mass-shell since the source emits
over a (wide) range of energies. A physical consequence is the possibility
that the particle {\it losses} its identity within the medium. In the
midsts of a high energy heavy-ion collision, such possibility can be realized
in the recombination of a jet parton with the partons from the surrounding
medium. Recall that during the propagation inside a deconfined QCD plasma, a 
fast parton can have not only induced gluon radiation but also {\it induced
absorption} from thermal gluons. This process can fairly well be considered as
parton recombination, which is one of the accepted mechanisms that are
used to describe the distinct features of meson and baryon spectra that
include a baryon to meson ratio larger than one for $p_t\gtrsim 2$ GeV in
central Au + Au collisions at RHIC~\cite{ppi} and their different azimuthal 
anisotropies~\cite{anisot}. This point is addressed in the context of the
modification of parton fragmentation functions induced by medium effects in
Ref.~\cite{Majumder}. When a parton recombines it certainly losses its
original identity and the energy loss should stop being described in terms of
parton degrees of freedom. Parton recombination from a jet with thermal
partons to form intermediate $p_t$ hadrons is a viable scenario in the case of
light flavors, given the features of the proton to pion ratio and even for $s$
quarks, given the features of the $\Lambda$ to kaon
ratio~\cite{klambda}. Other off mass-shell effects can be of relevance as well 
when studying if and how the virtuality of the propagating parton affects the
in-medium splitting functions~\cite{Wiedemann}.

In this work we study one of such off mass-shell effects, namely a possible
finite {\it life-time} of the scattered partons --originating within and
traversing the finite size medium-- in the description of the collisional
energy loss mechanism. We introduce the possibility that the scattered parton
be described in terms of a propagator containing a parameter associated to
the parton's life time. We argue that for recombining partons, this picture
could be used to consider collisional energy losses only up to times when
these recombine with thermal partons from the medium from where the energy
loss process should start being described in terms of hadronic degrees of
freedom.

The work is organized as follows: In Sec.~\ref{sec2} we rewrite the expression
for the collisional energy loss in a finite size QCD medium based on the
formalism used in Ref.~\cite{Djordjevic}, allowing for particles being emitted
off mass-shell and having a finite life-time. In Sec.~\ref{sec3} we give the
numerical estimates for the collisional energy losses of heavy and light
quarks using parameters relevant for RHIC energies. We use the cases studied
in Ref.~\cite{Djordjevic} as a base to compare our results to. We finally
conclude and give an outlook of the consequences of the result in
Sec.~\ref{sec4}.

\section{Energy loss}\label{sec2}

\begin{figure}[t] 
\vspace{-1cm}
\hspace{-15mm}
{\includegraphics[height=3.8in]{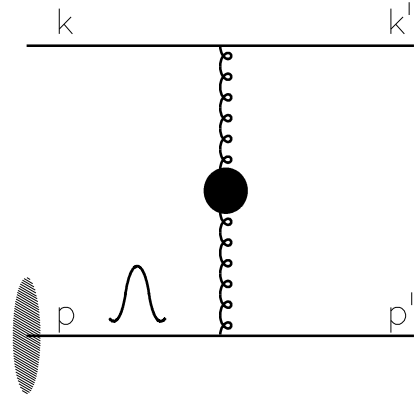}}
\vspace{-2cm}
\caption{Scattering diagram describing the lowest order
contribution to the energy loss due to the collision of fermions with incoming
and outgoing momenta $P$ and $P'$ with medium fermions with initial and final
momenta $K$ and $K'$, respectively. The blob in the intermediate gluon line
represents the effective HTL propagator. Incoming fermions are produced by the
source in wave packets and not necessarily on their mass-shell.}
\label{fig1}
\end{figure}

We start by describing the elemental interaction of a fermion with momentum
$P^\mu =(p_0,{\bf p})$ (not necessarily on its mass-shell), magnitude of its
velocity $v\equiv |{\bf v}|=p/E$,
mass $M$ and spin $s$ with a massless fermion in the non-expanding medium, with
momentum $K^\mu =(k,{\bf k})$ and spin $\lambda$, through the exchange of a
gauge boson with momentum $q^\mu=(\omega,{\bf q})$ and by means of a coupling
constant $g$. For elastic collisions, these particles retain their identities
and after the scattering they have momenta ${P'}^\mu =(E',{\bf p}')$ and
${K'}^\mu =(k',{\bf k}')$ and spins $s'$ and $\lambda '$, respectively, and
$E'=\sqrt{p'^2 + M^2}$. The scattering diagram associated to the process is
depicted in fig.~\ref{fig1}. When the incoming massive fermion is on its
mass-shell, i.e. $p_0=E=\sqrt{p^2 + M^2}$, the expression for the matrix
element describing this process is given by (see Eq.~(6) in
Ref.~\cite{Djordjevic})  
\bea
   i{\mathcal M}&=&-g^2\int d^4x\ j (t,{\bf x})e^{iP\cdot x}\int d^4x_1\int
   d^4x_2\nonumber\\ 
   &\times&\int\frac{d^3p}{(2\pi)^3 2E}
   \int\frac{d^4q}{(2\pi)^4}
   D_{\alpha\beta}(q)e^{iq\cdot (x_1-x_2)}\nonumber\\
   &\times&\bar{u}(p',s')e^{iP'\cdot x_1}\gamma^\alpha 
   u(p,s)e^{-iP\cdot x_1}\nonumber\\
   &\times&\bar{u}(k',\lambda ')e^{iK'\cdot x_2}\gamma^\beta
   u(k,\lambda)e^{-iK\cdot x_2}\nonumber\\
   &\times&\theta (t_1-t)\ \theta (L/v-(t_1-t)),
\label{matrixelement}
\eea
where $D_{\alpha\beta}$ will become the effective HTL gluon propagator, 
$j(t,{\bf x})e^{iP\cdot x}$ is the amplitude associated to the source to
produce an incoming particle with momentum $P$ and   
\bea
   x^\mu &=& (t,{\bf x}),\nonumber\\
   x^\mu_i&=&(t_i,{\bf x}_i),\hspace{0.5cm}i=1,2.
\label{defsvecs}
\eea
The step functions in Eq.~(\ref{matrixelement}) represent the conditions
that the produced fermion interacts within the plasma at $t_1$ after
being produced at $t$ and before it leaves the plasma at $L/v+t$, under
the approximation that its velocity remains constant.

When the source produces particles off their mass-shell, the matrix element
describing the process can be written as
\bea
   i{\mathcal M}&=&-\frac{g^2}{2M}\int d^4x\ j (t,{\bf x})\int d^4x_1\int
   d^4x_2\nonumber\\ 
   &\times&\int\frac{d^4p}{(2\pi)^4}
   \int\frac{d^4q}{(2\pi)^4}
   D_{\alpha\beta}(q)e^{iq\cdot (x_1-x_2)}\nonumber\\
   &\times&\bar{u}(p',s')e^{iP'\cdot x_1}\gamma^\alpha 
   S(P)e^{iP\cdot (x-x_1)}u(p,s)\nonumber\\
   &\times&\bar{u}(k',\lambda ')e^{iK'\cdot x_2}\gamma^\beta
   u(k,\lambda)e^{-iK\cdot x_2}\nonumber\\
   &\times&\theta (t_1-t)\ \theta (L/v-(t_1-t)),
\label{matrixelementmod}
\eea
where $S(P)e^{iP\cdot (x-x_1)}u(p,s)$ represents the amplitude to propagate a
fermion mode with momentum $P$ from $x$ to $x_1$. 
Notice that Eq.~(\ref{matrixelementmod}) reduces to Eq.~(\ref{matrixelement})
when $S(P)\rightarrow S_0(P)=(2\pi)(2M)\Lambda_+(P)\delta (P^2-M^2)\theta
(p_0)$, where  
\bea
   \Lambda_+(P)&=&\sum_{s}u(p,s)\bar{u}(p,s)\nonumber\\
   &=&\frac{\slsh{P} + M}{2M},
\label{lambda}
\eea
is the projector for positive-energy solutions.

Let us however consider the situation where $S(P)$ does not describe free,
on-mass shell propagation, but instead the propagation of a wave packet with a
finite width $\eta$. In order to explore a simple scenario, let us recall that
\be
   \lim_{\eta\rightarrow 0^+}
   \frac{\eta}{(p_0-E_p)^2+\eta^2}
   =(2E_p\pi)\delta (P^2-M^2)\theta (p_0),
\label{lim}
\ee
 where $E_p=+\sqrt{p^2+M^2}$. To consider a finite width, we take $\eta$
 finite and write
\be
   S(P)=\left(\frac{2M}{E_p}\right)\frac{\eta}{(p_0-E_p)^2+\eta^2}\Lambda_+(P).
\label{finiteta}
\ee
Upon the change of variable $y=x_1-x$ and after integration over $d^4x_2$,
$d^4y$, $d^3p$, $d^4q$ and $d^4x$ in Eq.~(\ref{matrixelementmod}) we get
\bea
   i{\mathcal M}&=&
   2i\left(\frac{\eta}{E_p}\right)\int_{-\infty}^\infty\frac{dp_0}{(2\pi)}
   e^{-i(p_0-E'-\omega)L/2v}\nonumber\\
   &\times&\frac{\sin [(p_0-E'-\omega)L/2v]}
   {(p_0-E'-\omega)[(p_0-E_p)^2+\eta^2]}\nonumber\\
   &\times&i{\mathcal M}_0(p_0)\ \tilde{j}(P),
\label{matrixelement2}
\eea
where $\tilde{j}$ is the Fourier transform of $j(t,\bf{x})$ and
\bea
   i{\mathcal M}_0(p_0)&=&ig^2
   D_{\alpha\beta}(K'-K)\nonumber\\
   &\times&\bar{u}(p',s')\gamma^\alpha 
   u(p,s)\bar{u}(k',\lambda ')\gamma^\beta
   u(k,\lambda)
\label{defs1}
\eea
and we used that $\Lambda_+(P)u(p,s)=u(p,s)$. ${\mathcal M}_0(p_0)$ is
the matrix element describing the scattering process in infinite volume and for
the following discussion, we have emphasized its dependence on $p_0$.

In order to perform the integral in Eq.~(\ref{matrixelement2}), we write 
\be
   \frac{1}{(p_0-E_p)^2+\eta^2}=
   \frac{1}{[(p_0-E_p)+i\eta][(p_0-E_p)-i\eta]},
\label{poles}
\ee
and for convergence, close the contour of integration on the
lower $p_0$ complex half-plane, which picks the pole at $p_0=E_p-i\eta$. This
results in the expression for the matrix element
\bea
   i{\mathcal M}&=&
   i\left(\frac{1}{E_p}\right)
   e^{-\eta L/2v}e^{-i(E_p-E'-\omega)L/2v}\nonumber\\
   &\times&\frac{\sin [(E_p-E'-\omega-i\eta )L/2v]}
   {(E_p-E'-\omega - i\eta)}\nonumber\\
   &\times&i{\mathcal M}_0(E_p-i\eta)\ \tilde{j}(P).
\label{aftermatrixelement2}
\eea
Notice that for a consistent description, we require the condition $\eta\ll
E_p$ meaning that the central energy of the wave packet is much larger than
it's width. Therfore in Eq.~(\ref{aftermatrixelement2}) we can approximate
\be
   {\mathcal M}_0(E_p-i\eta)\simeq{\mathcal M}_0(E_p).
\label{approx1}
\ee
This approximation cannot be made for the rest of the factors in
Eq.~(\ref{aftermatrixelement2}) since the term $E_p-E'-\omega$ is of order of
the transfered momentum that in turn is of order of the medium's temperature
which may not be much larger than the wave packet's width.

The square of the matrix element given in Eq.~(\ref{aftermatrixelement2}),
averaged over the initial spin $s$ and summed over all other spins is
therefore  
\bea
   \frac{1}{2}\sum_{s,s',\lambda ,\lambda '}|{\mathcal M}|^2
   &=&e^{-\eta L/v}\left(\frac{1}{E_p^2}\right)
   |\tilde{j}(P)|^2\nonumber\\
   &\times&\left|\frac{\sin[(\omega-{\bf v}\cdot{\bf q}+i\eta)L/2v]}
   {\omega-{\bf v}\cdot{\bf q}+i\eta}\right|^2\nonumber\\
   &\times&\frac{1}{2}\sum_{s,s',\lambda ,\lambda '}|{\mathcal M}_0(E_p)|^2,
\label{matrixelementsquared}
\eea 
where we use that for an energetic incoming fermion
$E_p-E'\simeq {\bf v}\cdot{\bf q}$. It is worth mentioning that if instead of
using the propagator in Eq.~(\ref{finiteta}), one uses the free Feynman
propagator, the result for the square of the matrix element, averaged over the
initial and summed over all other spins, yields the result found in 
Ref.~\cite{Djordjevic}.

Hereafter, we specialize to the description of the scattering of the fermion
(quark) in the QCD plasma. The differential energy loss $dE$ is related to the
differential collisional interaction rate $d\Gamma$ by $dE=\omega
d\Gamma$~\cite{Braaten1}. On the other hand, $d\Gamma$ is given by 
\bea
   d^3N d\Gamma &=& \frac{1}{2}\sum_{s,s',\lambda ,\lambda '}
   |{\mathcal M}|^2
   \frac{d^3p'}{(2\pi)^32E'}\frac{d^3k}{(2\pi)^32k}\frac{d^3k'}{(2\pi)^32k'}
   \nonumber\\
   &\times&
   \sum_{\xi=q,\ \bar{q},\ g}n^\xi_{eq}(k)[1\pm n^\xi_{eq}(k')]\nonumber\\
   &\simeq& \frac{1}{2}\sum_{s,s',\lambda ,\lambda '}
   |{\mathcal M}|^2
   \frac{d^3p'}{(2\pi)^32E'}\frac{d^3k}{(2\pi)^32k}\frac{d^3k'}{(2\pi)^32k'}
  \nonumber\\
   &\times& n_{eq}(k),
\label{diffrate}
\eea
where 
\bea
   d^3N&=&d_R|j(P')|^2[u(p',s')\bar{u}(p',s')]^2\frac{d^3p'}{(2\pi)^32E'}
   \nonumber\\
   &=&d_R|j(P')|^2\frac{d^3p'}{(2\pi)^32E'},
\label{incnumb}
\eea
represents the number of (non-scattered) particles into the phase space volume
in the interval ${\bf p}'$ and ${\bf p}' + d^3p'$ with $d_R=3$ for 
the $SU(3)$ fundamental representation. Also, in Eq.~(\ref{diffrate}) we have
used that when describing the collisional energy loss, 
$n^\xi_{eq}(k)[1\pm n^\xi_{eq}(k')]= n^\xi_{eq}(k)$ since the term 
proportional to $n^\xi_{eq}(k)n^\xi_{eq}(k')$ is odd under the exchange of $k$
and $k'$ and integrates to zero~\cite{Braaten1}, and have defined
\be
   n_{eq}(k)=\sum_{\xi=q,\ \bar{q},\ g}n^\xi_{eq}(k).
\label{defeqdist}
\ee
\begin{figure}[t] 
{\includegraphics[height=2.1in]{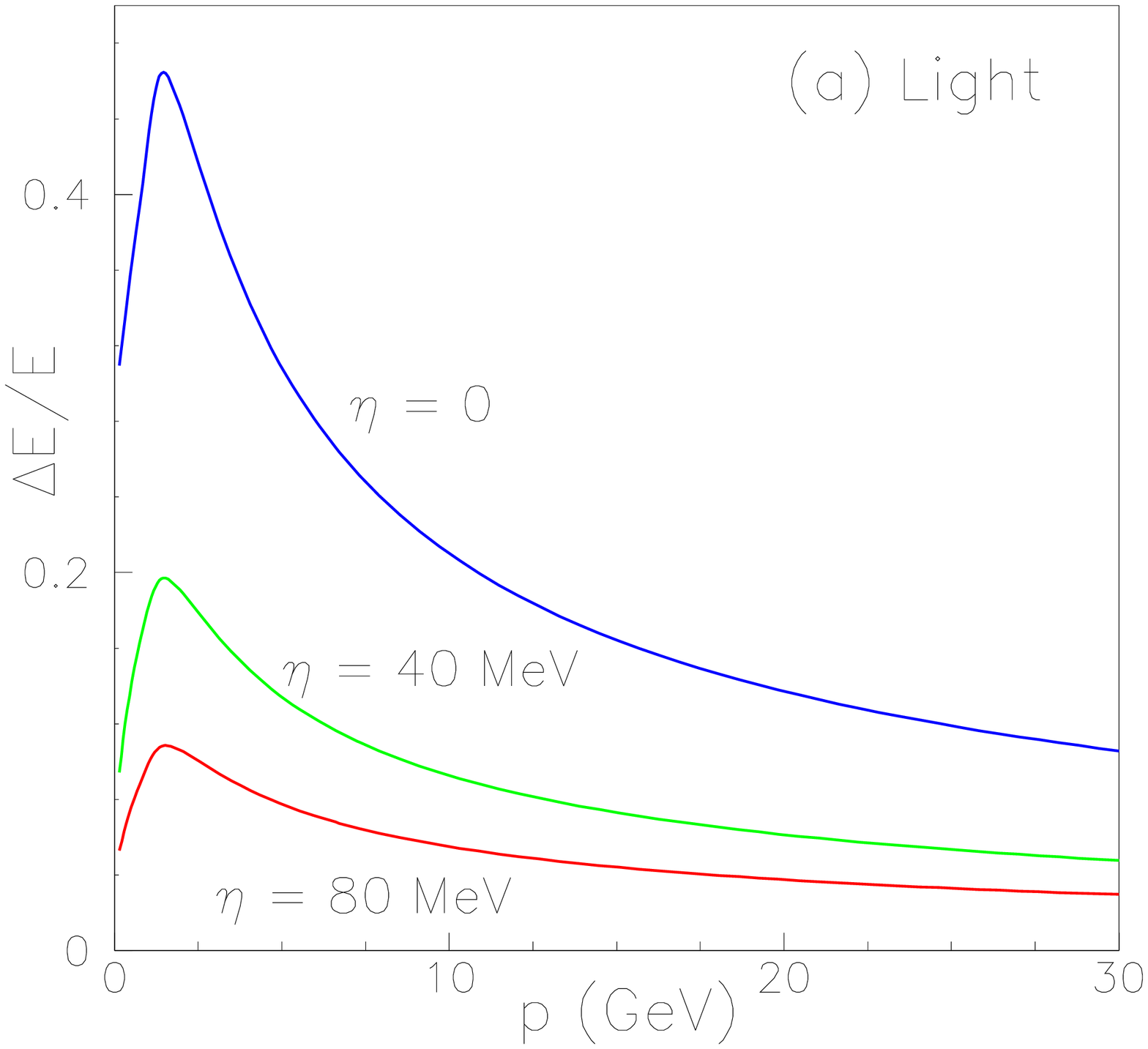}
 \includegraphics[height=2.1in]{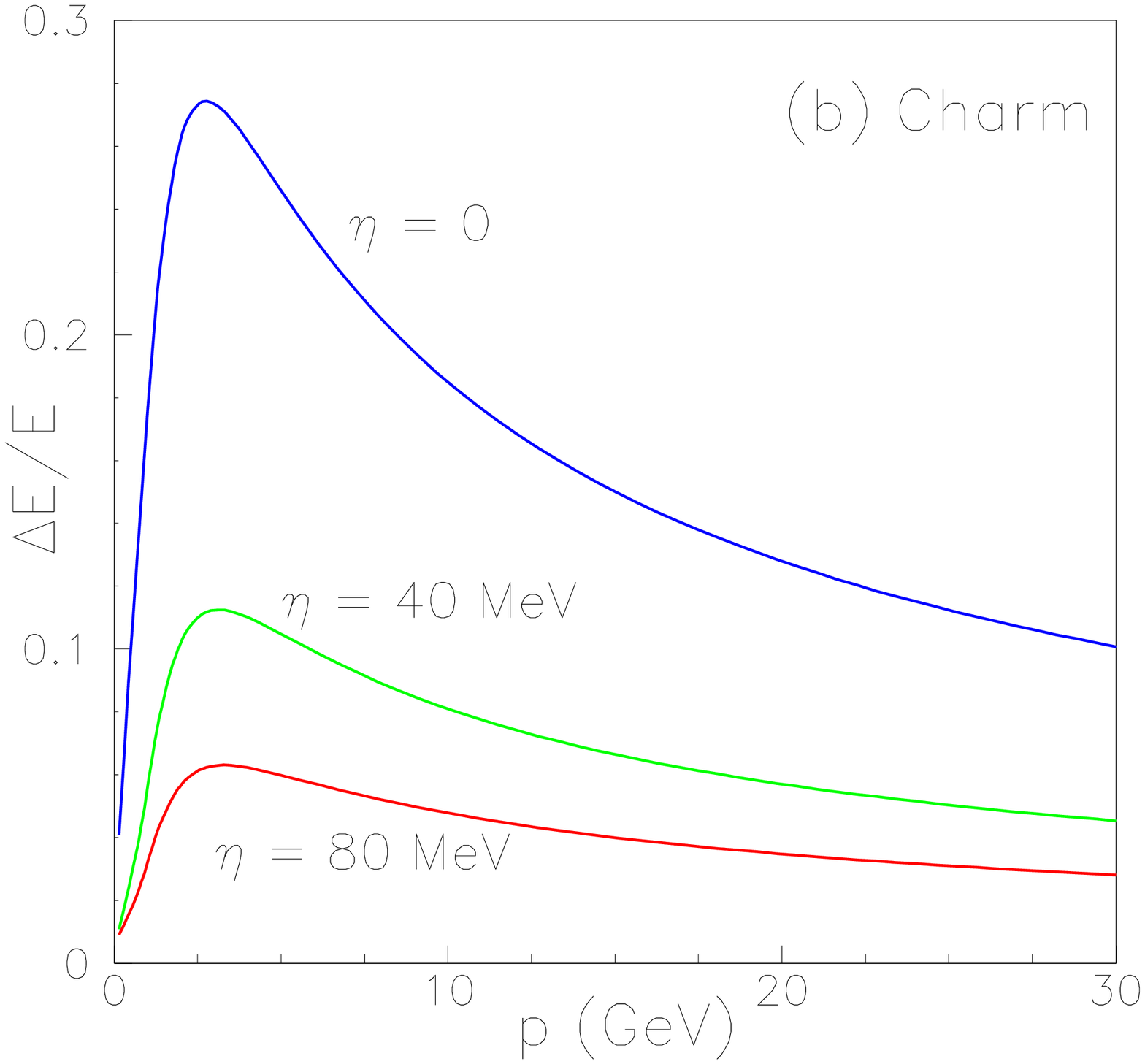}
 \includegraphics[height=2.1in]{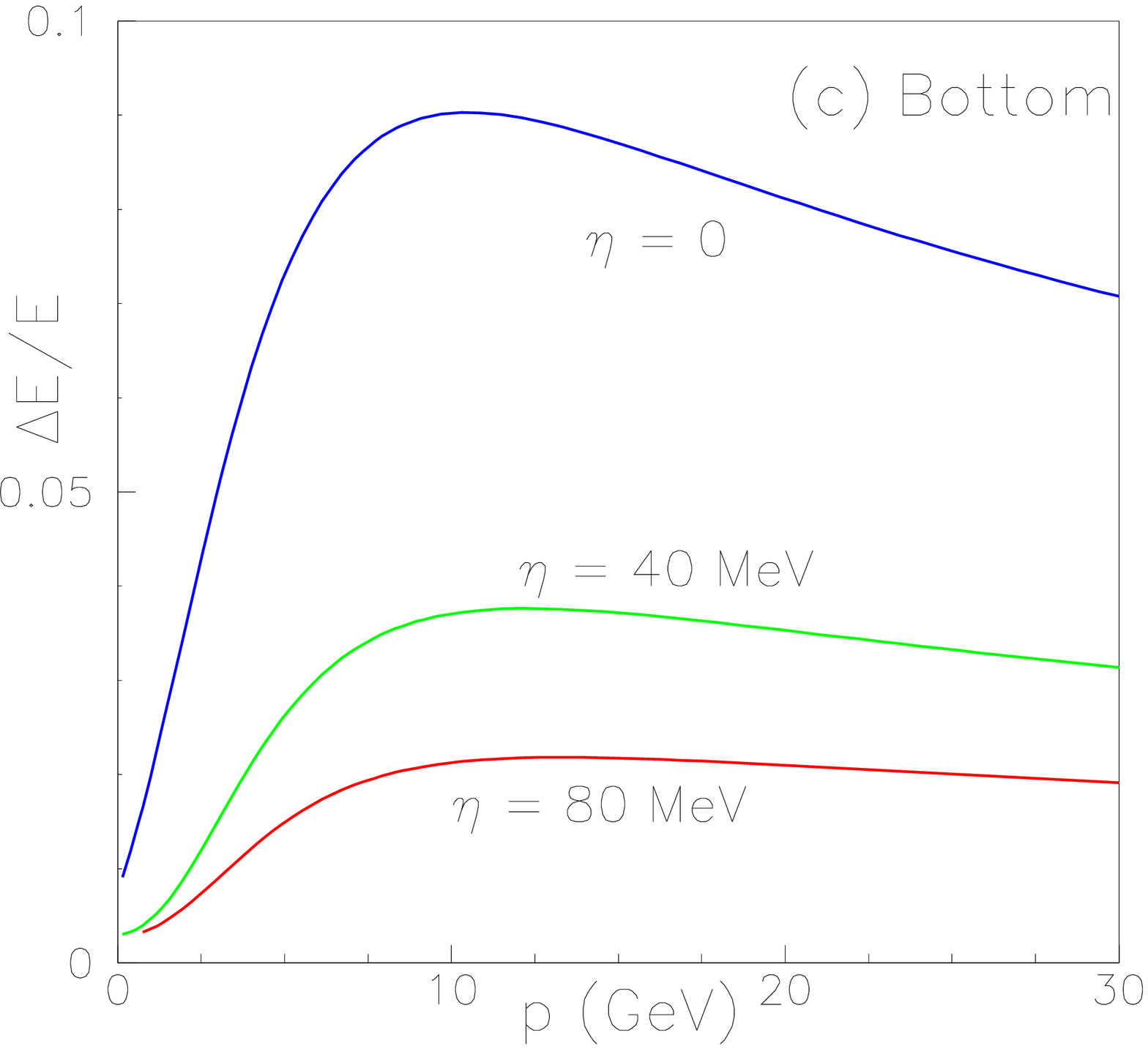}
}
\caption{(Color online) Fractional energy loss for (a) light, (b) charm and
 (c) bottom quarks as a function of their momenta for a fixed medium's length
 $L=5$ fm. The uppermost curve in each
 case corresponds to the description without off mass-shell effects. This is
 compared to the case with off mass-shell effects for two values of $\eta=$
 40, 80 MeV. In each case, the fractional energy loss decreases as the value
 of $\eta$ increases.} 
\label{fig2}
\end{figure}
For a source producing energetic particles with a large spread in momentum, we
can take the approximation $|j(P')|^2\simeq |j(P)|^2$. Therefore, the
collisional energy loss, considering the finite width of the scattered wave
packet can be written as
\bea
   \Delta E&\simeq& C_R\frac{e^{-\eta L/v}}{E_p^2}
   \int\frac{d^3k}{(2\pi)^32k}n_{eq}(k)\int\frac{d^3k'}{(2\pi)^32k'}\omega
   \nonumber\\
   &\times&\left|\frac{\sin[(\omega-{\bf v}\cdot{\bf q}+i\eta)L/2v]}
   {\omega-{\bf v}\cdot{\bf q}+i\eta}\right|^2\nonumber\\
   &\times&\frac{1}{2}\sum_{s,s',\lambda ,\lambda '}|{\mathcal M}_0(E_p)|^2.
\label{energyloss}
\eea
Notice that Eq.~(\ref{energyloss}) is modified with respect to the
corresponding expression in Ref.~\cite{Djordjevic} by the $\eta$-dependent
exponential factor and the $\eta$ dependence in the arguments of the sine
function and in the energy denominator. When $\eta\rightarrow 0$, the
corresponding expression for the energy loss in Ref.~\cite{Djordjevic} is
recovered. In order to find the explicit expression for
Eq.~(\ref{energyloss}), recall that the effective gluon propagator can be
written as
\be
   D^{\mu\nu}=-P^{\mu\nu}\Delta_T-QP^{\mu\nu}\Delta_L
\label{gluoneff}
\ee
where in the HTL approximation, the effective transverse and longitudinal gluon
propagators are given by
\bea
   \Delta_T^{-1}&=&\omega^2-q^2-\frac{m_D^2}{2}-
   \frac{(\omega^2-q^2)m_D^2}{2q^2}\nonumber\\
   &\times&
   \left( 1 + \frac{\omega}{2q}
   \ln \left|\frac{\omega - q}{\omega + q}\right|\right)\nonumber\\
   \Delta_L^{-1}&=&q^2+m_D^2\left(1+\frac{\omega}{2q}
   \ln \left|\frac{\omega - q}{\omega + q}\right|\right),
\label{gluoneffTL}
\eea
where $m_D^2=g^2T^2(1+N_f/6)$ is the square of the Debye mass and, working in
Coulomb gauge, the only non-vanishing components of the transverse and
longitudinal projectors are 
\bea
   P^{ij}&=&\delta^{ij} - \frac{q^iq^j}{q^2}\nonumber\\
   Q^{00}&=&1.
\label{projCou}
\eea
\begin{figure}[t] 
{\includegraphics[height=2.1in]{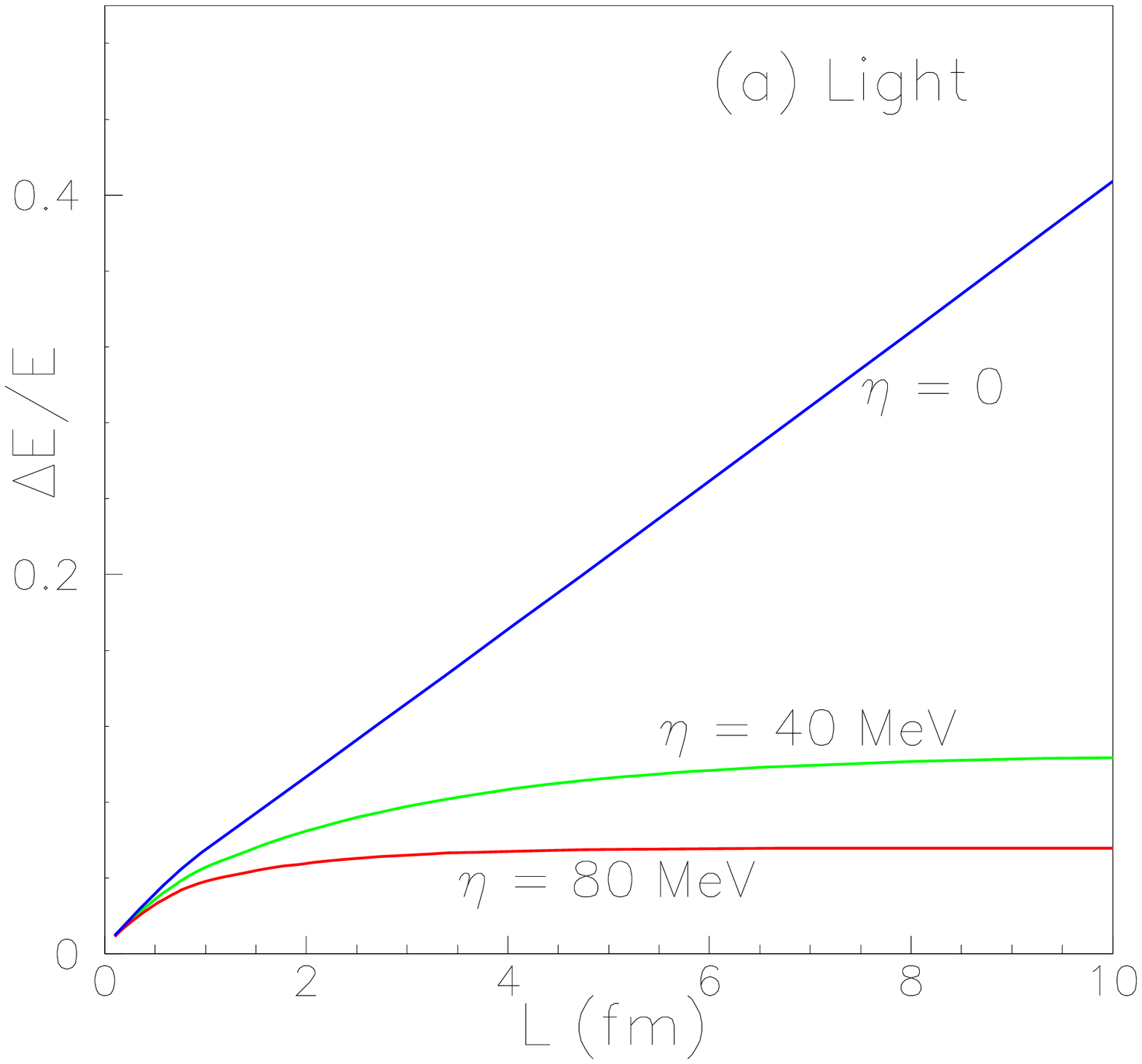}
 \includegraphics[height=2.1in]{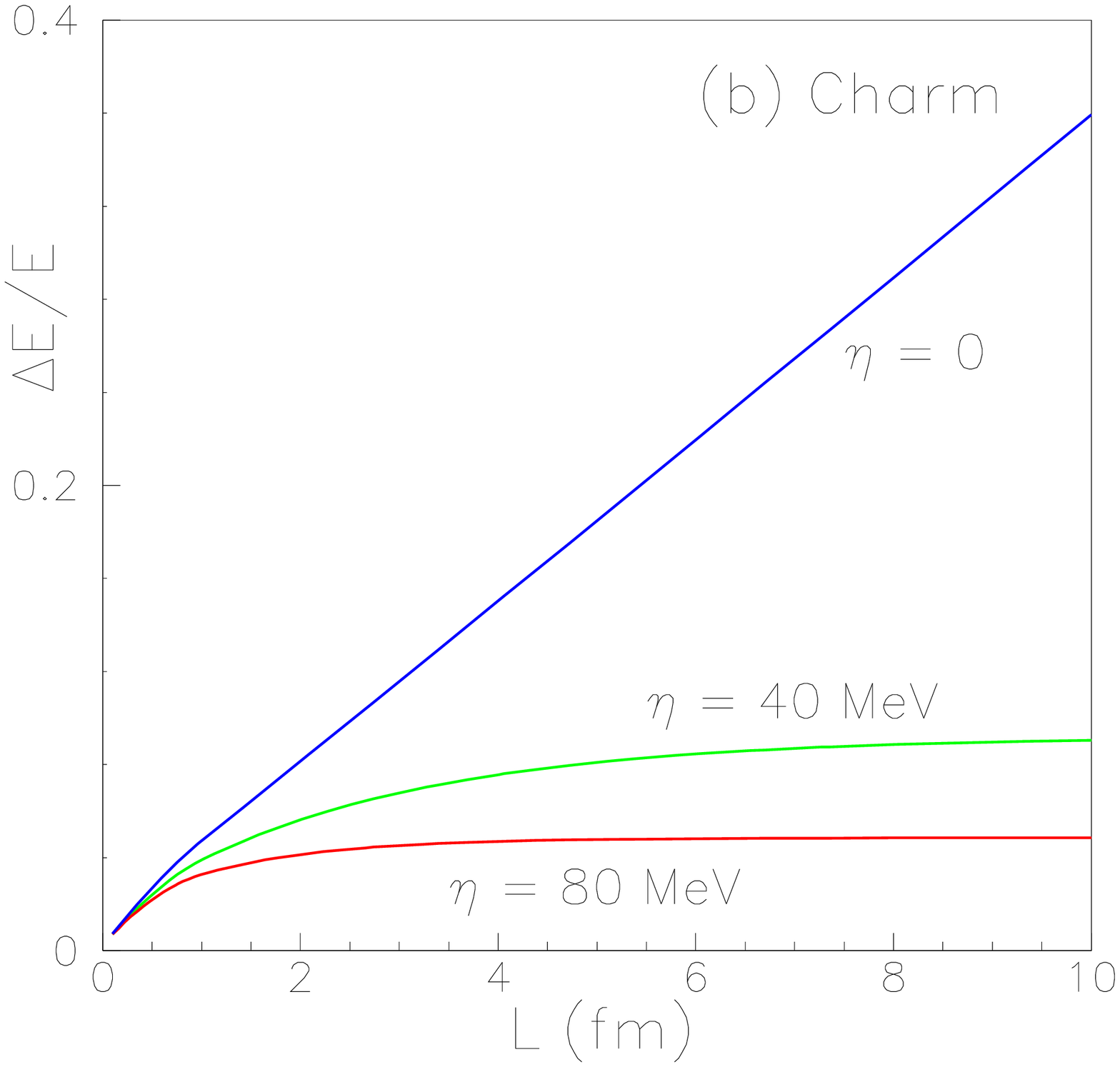}
 \includegraphics[height=2.1in]{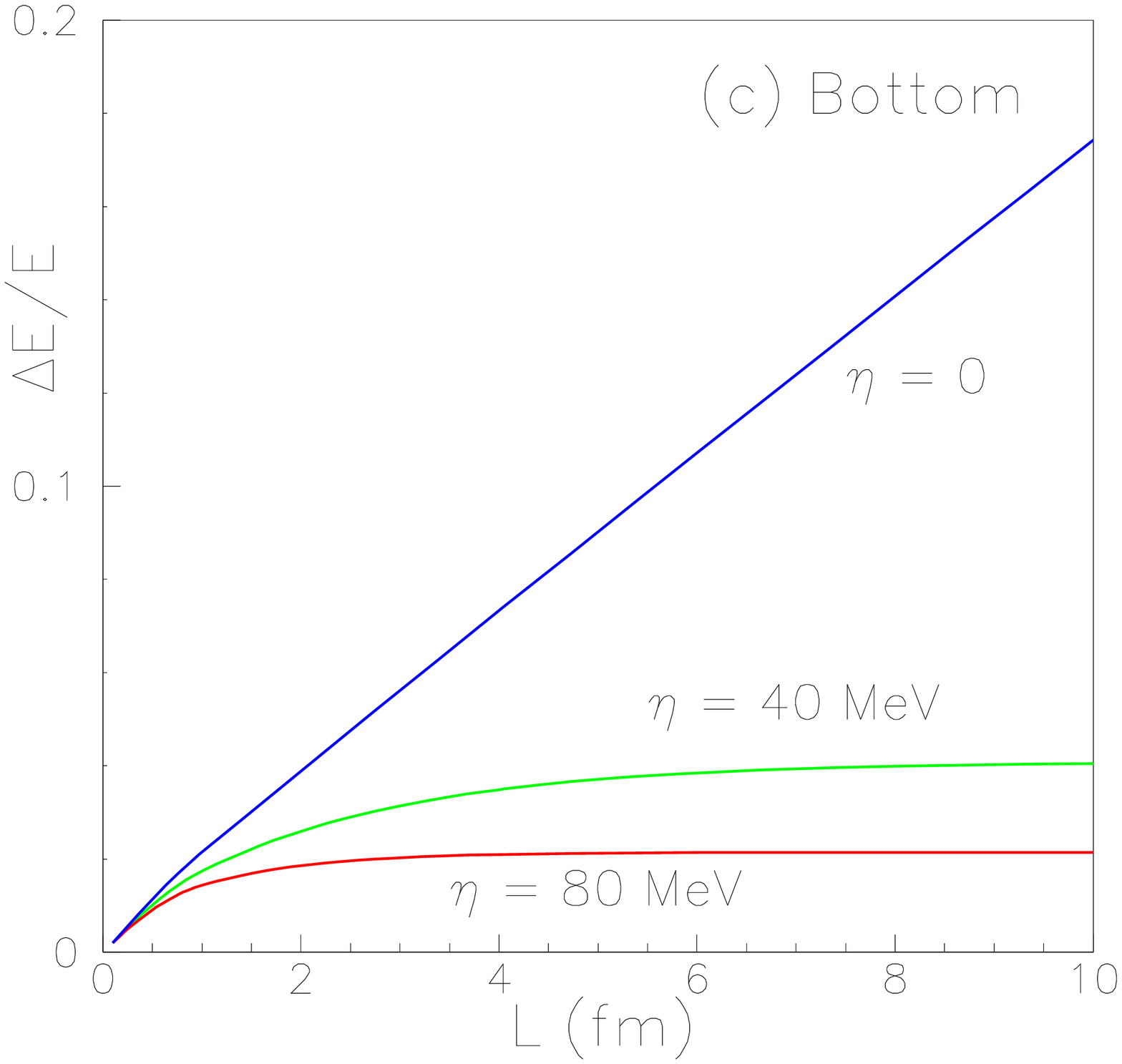}
}
\caption{(Color online) Fractional energy loss for (a) light, (b) charm and
 (c) bottom quarks as a function of the medium's length for a fixed quark
 momentum  $p=10$ GeV. The uppermost curve in each case corresponds to 
 the description without off mass-shell effects. This is compared to the case
 with off mass-shell effects for two values of $\eta=$ 40, 80 MeV. In each
 case, the fractional energy loss decreases as the value of $\eta$ increases.}
\label{fig3}
\end{figure}
Using Eqs.~(\ref{gluoneff})~--~(\ref{projCou}), the  averaged over
initial, summed over all other spins, matrix element squared describing the
underlying scattering process in vacuum is given by
\bea
   \frac{1}{2}\sum_{s,s',\lambda ,\lambda'}|{\mathcal M}_0|^{2}
   &=& 16g^{4}E^{2}_p\Biggl\{     
   | \Delta_{L}(q) |^{2}\left( kk' + {\bf k}\cdot{\bf k}' \right)\nonumber\\
   &+&
   2\ {\mbox {Re}}\left[\Delta(q)_L\Delta(q)_T^{*} \right]\nonumber\\
   &\times&\Biggl[ 
   k\left(\vv\cdot\kvp
   -\frac{\vv\cdot\qv\ \kvp\cdot\qv}{q^{2}}\right)\nonumber\\
   &+& k'\left(\vv\cdot\kv
   -\frac{\vv\cdot\qv\ \kv\cdot\qv}{q^{2}}\right) 
   \Biggr]\nonumber\\
   &+&| \Delta_T(q) |^{2}\Biggl[2\left(\vv\cdot\kv
   -\frac{\vv\cdot\qv\ \kv\cdot\qv}{q^{2}}\right) \nonumber\\
   &\times&
   \left(\vv\cdot\kvp
   -\frac{\vv\cdot\qv\ \kvp\cdot\qv}{q^{2}}\right)\nonumber\\
   &+&\left(kk' - \kv\cdot\kvp \right)
   \left(v^2-\frac{\vv\cdot\qv\ \vv\cdot\qv}{q^{2}}\right) 
   \Biggr]
   \Biggr\}.\nonumber\\
\label{matelzero}
\eea
Since for a non-expanding medium, the energy loss does not depend on the
direction of ${\bf v}$, we can simplify Eq.~(\ref{energyloss}) by averaging
over the direction of ${\bf v}$~\cite{Braaten1}. This is most conveniently
performed by introducing the auxiliary functions
\be
   {\mathcal J}_i=\int\frac{d\Omega}{4\pi}
   \left|\frac{\sin[(\omega-{\bf v}\cdot{\bf q}+i\eta)L/2v]}
   {\omega-{\bf v}\cdot{\bf q}+i\eta}\right|^2
   (\omega-{\bf v}\cdot{\bf q})^i,
\label{Js}
\ee
$i=1,\ 2,\ 3$, in terms of which, the average over the different powers of
${\bf v}$ appearing in Eq.~(\ref{matelzero}) can be expressed. The functions
in Eq.~(\ref{Js}) are explicitly given in the appendix. After averaging over
the directions of ${\bf v}$, the expression for the energy loss can be written
as 
\bea
   \Delta E&=&\frac{C_Rg^4}{2\pi^4}e^{-\eta L/v}
   \int_0^\infty n_{eq}(k)dk\nonumber\\
   &\times&\left(\int_0^kqdq\int_{-q}^q\omega d\omega
   + \int_k^{q_{\mbox{\tiny{max}}}}qdq\int_{q-2k}^q\omega d\omega\right)
   \nonumber\\
   &\times&
   \Bigg(|\Delta_L(q)|^2\frac{(2k+\omega)^2-q^2}{2}{\mathcal J}_0
   \nonumber\\
   &+&|\Delta_T|^2\frac{[q^2-\omega^2][(2k+\omega)^2+q^2]}{4q^4}\nonumber\\
   &\times&[(v^2q^2 - \omega^2){\mathcal J}_0 + 2\omega {\mathcal J}_1
   -{\mathcal J}_2]\Bigg),
\label{energylossexpl}
\eea
where the limits of integration over $\omega$ and $q$ take into account that
for the considered scattering amplitude, the transferred four-momentum is
space-like and
\be
   q_{\mbox{\tiny{max}}}=\frac{2k(1+k/E_p)}{1-v+2k/E_p},
\label{qmax}
\ee
is obtained from the approximation that the maximum energy transfered occurs
for backward scattering~\cite{Thoma}.

\section{Numerical results}\label{sec3}

In order to present the quantitative behavior for the energy loss, we take
standard values for the parameters involved. We give examples of the effect
for light as well as for heavy flavors both, to study the mass effect and to
directly compare to the findings of Ref.~\cite{Djordjevic}. The plasma
temperature is taken as $T=0.225$ GeV, the effective number of flavors
$N_f=2.5$, the strength of the coupling constant $\alpha = g^2/4\pi=0.3$ and
the Debye mass $m_D=0.5$ GeV. The bottom quark mass is taken as 4.5 GeV
whereas the charm quark mass is taken as 1.2 GeV. We take the mass of the
light quarks as 0.2 GeV.

Figure~\ref{fig2} shows the fractional energy loss $\Delta E/E$ for light,
charm and bottom quarks in a finite size medium with $L=5$ fm, for the cases
with and without off mass-shell effects. For the curves describing the off
mass-shell effects, we consider two values, $\eta=40$ MeV and $\eta=80$
MeV. Notice that in all cases, a finite value of $\eta$ produces the 
energy loss to decrease as compared to the case $\eta =0$. The decrease is
more important for larger values of $\eta$. 

Figure~\ref{fig3} shows the fractional energy loss for light, charm and
bottom quarks as a function of the medium's length $L$ for a fixed quark
momentum $p=10$ GeV, comparing also the cases with and without off mass-shell
effects. For the curves describing the off mass-shell effects, we consider
once more the two values, $\eta=40$ MeV and $\eta=80$ MeV. Notice that a
finite value of $\eta$ causes the fractional energy loss to asymptotically
reach a maximum value as the medium's size increases. This is in sharp
contrast with the case where no off mass-shell effects are considered where
for large $L$ the energy loss increases linearly.

\begin{figure}[t] 
\hspace{45mm}
{\centering
{\includegraphics[height=2.8in]{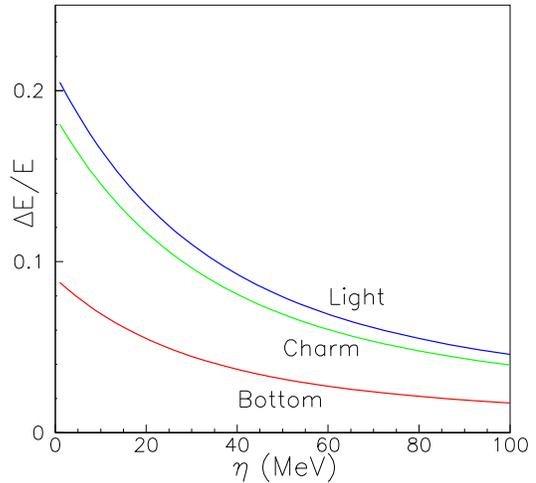}
}}
\caption{(Color online) Fractional collisional energy loss for light, charm
 and bottom quarks as a function of $\eta$ for a fixed quark momentum $p=10$
 GeV and a fixed medium's size $L=5$ fm. The fractional energy loss decreases
 with increasing $\eta$ and the decrease is similar in shape independent of
 the quark mass.} 
\label{fig4}
\end{figure}

Figure~\ref{fig4} shows the behavior of the fractional energy loss as a
function of $\eta$ also for light, charm and bottom quarks for a fixed value
of the quark momentum $p=10$ GeV and a fixed value of the medium's size $L=5$
fm. The fractional energy loss decreases with increasing $\eta$ and the
decrease is similar in shape, regardless of the quark mass.

\section{Discussion and conclusions}\label{sec4}

In this work we have studied the off mass-shell effects on the collisional
energy loss of particles, produced and scattered within a finite size QCD
medium, associated to the introduction of a finite-width wave packet and
therefore a finite particle's life-time. We have shown that this effect
decreases the energy loss as compared to the case when these particles are
produced on mass-shell and therefore live longer than the medium, fragmenting
outside it. We have argued that this picture should be applied in particular
to energetic partons that recombine with thermal partons and thus hadronize
within the medium.

Recall that the length scales playing a role for the energy loss mechanisms in
a finite size, thermal, non-expanding medium are the medium's size 
$L$, the average distance between collisions $d\sim 1/T$, the Debye radius
$r_D$, the mean free path $\delta\sim 1/g^2 T$ and the particle's formation
time $t_f\sim 1/E_p$. When considering the in-medium particle's lifetime, the
length scale $\eta^{-1}$ is introduced. For the description of the
scattering process in terms of a perturbative picture, it is required that the
hierarchy of scales
\be
   1/T\ll r_D\ll \delta
\label{hirarchy}
\ee
be satisfied. The requirement that the scattering particle can be described in
terms of an oscillating mode means that this mode is not damped too fast,
which in turn translates into the condition 
\be
   1/E_p\ll 1/\eta,
\label{condition}
\ee
and can be thought of as the condition that the time the medium takes to
produce the particle is much shorter than the in-medium particle's
lifetime. For very energetic partons, it is safe to assume that $t_f\sim
1/E_p$ is the smallest of all length scales. Moreover, for media sizes of the
order expected to be produced in relativistic heavy-ion collisions, it is also
safe to assume that $L$ is larger than the mean free path and therefore the
hierarchy of scales
\be
   1/E_p\ll 1/T\ll r_D\ll \delta \lesssim L,
\label{hierarchy2}
\ee
follows. In fact, the results in Refs.~\cite{Djordjevic,Gossiaux} can be
viewed as meaning that as long as the medium size is larger than the Debye
radius, the collisional energy loss for on mass-shell particles in a finite
size medium is not suppressed as compared to the infinite medium case. 

The situation changes when introducing the in-medium particles' life-time. The
results of this work show that when $\eta^{-1}\lesssim L$ the effects are
strong. They cease to matter for $\eta\rightarrow 0$. 

It should be emphasized that in the context of this work, the term {\it loss
of identity} does not refer to a parton change of color or phase, which are
accounted for already in the description of the underlying QCD scattering
process, neither to a change of flavor which would be mediated by the weak
interaction and thus irrelevant for the time scales involved during the QCD
plasma phase. This term is rather related to the onset of a hadronization
mechanism that happens during the interaction of a fast (hard) parton with
soft ones from the medium in such a way as to produce a hadron by means of
recombination. This process has been referred to as {\it shower and thermal
parton recombination}, in Ref.~\cite{Hwa}. In this way, the loss of identity
is related to the fact that when this kind of partons form a hadron, the
energy loss should stop being described in terms of parton degrees of freedom
and start being described in terms of hadron degrees of freedom, thereby
effectively producing the parton to disappear from the description.  

Typical time scales involved in hard-soft parton recombination are of
order $\tau_{\mbox{\tiny{recomb}}} \sim 1.5$ fm, despite the low momentum
transfers involved, since, as argued for instance in Ref.~\cite{Casalderrey},
this recombination needs not be local and it can be mediated by a QCD
string. Such time scale is well within the life-time of the QCD medium, for
central collisions and the largest nuclei, where it is estimated to be of
order 5 fm.

When the medium length shortens and falls below $\tau_{\mbox{\tiny{recomb}}}$,
hadronization is more likely to happen outside the medium. This means in
particular that for peripheral collisions or collisions of smaller systems,
the energy loss description in terms of partonic degrees of freedom is
appropriate. This is accounted for in our description when we take $\eta^{-1} >
L$, for which the energy loss with and without the use of the parameter $\eta$
coincide. Indeed, a change in the energy loss of intermediate momentum hadrons
should exist as a function of centrality and system size. In order to quantify
such change it is necessary to quantitatively estimate the energy loss of a
hadron within a QCD medium. This calculation is for the moment outside the
scope of the present work.  

In the context of recombination, the use of the parameter $\eta$ should be
introduced into a statistical scenario that also incorporates the evolution of
the colliding system with energy density, as well as using a realistic
geometry, including adequate probability profiles to produce jets~\cite{Paic}
and the effects of an expanding medium. All this is for future.

\section*{Acknowledgments}

A.A. thanks N. Armesto for very useful conversations during the genesis of
this work and his kind hospitality during a visit to the U. of Santiago de
Compostela in the summer of 2006. Support has been received in part by a HELEN
grant, PAPIIT-UNAM grant number IN107105, CONACyT grant number 40025-F,
bilateral agreement CONACyT-CNPq grant numbers J200.556/2004 and
491227/2004-3 and FAPERJ (Brazil) under contract Project No. E-26/170.158/2005.

\section{Appendix}

The functions defined in Eq.~(\ref{Js}) are explicitly given by 
\begin{widetext}
\bea
   {\mathcal J}_0&\equiv&\int\frac{d\Omega}{4\pi}
   \left|\frac{\sin[(\omega-{\bf v}\cdot{\bf q}+i\eta)L/2v]}
   {\omega-{\bf v}\cdot{\bf q}+i\eta}\right|^2\nonumber\\
   &=&-\frac{1}{8qv\eta}\left[
   i\cosh (\eta L/v)
   \left(\sum_{l,l'=\pm 1}
   {\mbox {sgn}} (l) {\mbox {sgn}} (l') Ci[(\omega-lqv-l'i\eta )L/v]\right.
   \right.
   \nonumber\\
   &+&2i\left.
        \left[\arctan \left(\frac{qv + \omega}{\eta}\right) +
              \arctan \left(\frac{qv - \omega}{\eta}\right)
        \right]
        \right)
   \nonumber\\
   &+&\sinh (\eta L/v)\left.
   \left(\sum_{l,l'=\pm 1}
   {\mbox {sgn}} (l) Si[(\omega-lqv-l'i\eta)L/v]\right)\right]\nonumber\\ 
   {\mathcal J}_1&\equiv&\int\frac{d\Omega}{4\pi}
   \left|\frac{\sin[(\omega-{\bf v}\cdot{\bf q}+i\eta)L/2v]}
   {\omega-{\bf v}\cdot{\bf q}+i\eta}\right|^2
   (\omega-{\bf v}\cdot{\bf q})\nonumber\\
   &=&\frac{1}{8qv}\left[
   \cosh (\eta L/v)
   \left(\sum_{l,l'=\pm 1}
   {\mbox {sgn}} (l) Ci[(\omega-lqv-l'i\eta )L/v]\right.\right.\nonumber\\
   &+&\left.\ln\left[
   \frac{(\omega + qv)^2+\eta^2}{(\omega - qv)^2+\eta^2}\right]\right)
   \nonumber\\
   &-&i\left.\sinh (\eta L/v)
   \left(\sum_{l,l'=\pm 1}
   {\mbox {sgn}} (l){\mbox {sgn}} (l') Si[(\omega-lqv-l'i\eta )L/v]\right)
   \right],
   \nonumber\\
   {\mathcal J}_2&\equiv&\int\frac{d\Omega}{4\pi}
   \left|\frac{\sin[(\omega-{\bf v}\cdot{\bf q}+i\eta)L/2v]}
   {\omega-{\bf v}\cdot{\bf q}+i\eta}\right|^2
   (\omega-{\bf v}\cdot{\bf q})^2\nonumber\\
   &=&\frac{1}{8qv}\left[
   \eta\cosh (\eta L/v)
   \left(i\sum_{l,l'=\pm 1}
   {\mbox {sgn}} (l){\mbox {sgn}} (l') 
   Ci[(\omega-lqv-l'i\eta )L/v]\right.\right.
   \nonumber\\
   &-&2\left.
        \left[\arctan \left(\frac{qv + \omega}{\eta}\right) +
              \arctan \left(\frac{qv - \omega}{\eta}\right)
        \right] + \frac{4qv}{\eta}\right)\nonumber\\
   &+& \eta\sinh (\eta L/v)
   \left(\sum_{l,l'=\pm 1}
   {\mbox {sgn}} (l) Si[(\omega-lqv-l'i\eta )L/v]\right)\nonumber\\
   &-&\left.\left(\frac{4v}{L}\right)\cos (L\omega /v)\sin (Lq)\right],
   \label{Jexplicit}
   \eea
\end{widetext}
where ${\mbox {sng}}$ is the sign function and $Ci$ and $Si$ are the cosine
and sine integrals, respectively. Despite their appearance, the above
functions are all real for real values of $\omega,\ q,\ v,\ L$ and $\eta$.

\end{document}